# Observation of two-level critical state in the superconducting FeTe thin films

Hao Ru(茹浩), Yi-Shi Lin(林一石), Yin-Cong Chen(陈寅聪), Yang Feng(冯洋), Yi-Hua Wang(王熠华)[*]

Department of Physics and State Key Laboratory of Surface Physics, Fudan University, Shanghai China 200438

[*] Correspondence author. Email: wangyhv@fudan.edu.cn

**FeTe, a non-superconducting parent compound in the iron-chalcogenide family, becomes superconducting after annealing in oxygen. Under the presence of magnetism, spin-orbit coupling, inhomogeneity and lattice distortion, the nature of its superconductivity is not well understood. Here, we combined mutual inductance technique with magneto transport to study the magnetization and superconductivity of FeTe thin films. We found that the films with the highest Tc showed non-saturating superfluid density and a strong magnetic hysteresis distinct from that in a homogeneous superconductor. Such hysteresis can be well explained by a two-level critical state model and suggested the importance of granularity to superconductivity in this compound.**

Iron-chalcogenide is an important family of iron-based high $T_C$ superconductors (FeSC). It has the simplest crystal structure with one layer of Fe square lattice sandwiched between two layers of chalcogen atoms. Despite the simplicity of its crystal structure, a single layer of FeSe grown on SrTiO$_3$ displays the highest T$_C$ among FeSC[1]. While the electronic structures are similar across FeSC, the magnetic structure in the Fe square lattice is complex and intriguing[2,3], and is believed to be essential for the pairing interaction in FeSC[4]. As spin-fluctuations on the FeSe side evolve into an antiferromagnetic ordering on the FeTe side, superconductivity disappears. Furthermore, with the presence of stronger spin-orbit coupling at higher Te concentration, the band structure of iron-chalcogenides has shown non-trivial topology[5] and the vortex core of the superconducting compounds have shown signs of zero-bias peaks[6]. Inducing superconductivity in FeTe is therefore useful both for the understanding of its relation with magnetism and for utilizing even stronger spin-orbit coupling in iron-chalcogenide family.

It has been reported that FeTe films may become superconducting after annealing in oxygen[7] or simply being exposed to air for a long time[8]. The oxygen atoms occupy the interstitial sites in the Te planes and substituting Te with O would only suppress superconductivity[9]. Exposing a bulk crystal to oxygen did not show similar effect[10], suggesting that the substrate plays a role in inducing superconductivity. Nevertheless, superconductivity has been found in FeTe films grown on different substrates [7, 8,10], even those that do not necessarily match the crystal symmetry and lattice constant of FeTe [7,10]. The exact roles of oxygen and substrate in inducing the superconductivity in FeTe and whether such superconductivity bears any resemblance to other compounds in the iron-chalcogenide family remained largely unknown.

In this letter, we report the observation of two-level critical state in the superconducting FeTe thin films grown on Al$_2$O$_3$. We combined mutual inductance technique with magneto transport to uncover the surface resistance change over 8 decades when the superconductivity was tuned by temperature and magnetic field. Despite a Tc of around 13 K, superfluid density did not saturate down to 2 K. Furthermore, the films showed magnetic hysteresis in surface resistance distinctively different from what one might expect from a homogeneous superconductor with vortex pinning. The hysteresis decreased with increasing temperature and reducing maximum magnetic field in a way consistent with the two-level critical state model. These observations suggested the importance of granularity for the superconductivity in FeTe thin films induced by oxygen.

We have grown FeTe thin films using molecular-beam-epitaxy on both $Al_2O_3$ and $SrTiO_3$ substrates and annealed the sample in-situ in an oxygen pressure of about $10^{-2}$ Torr (see supplementary materials). Films on the latter showed better morphology (see supplementary materials) but much lower or even zero Tc. The current study will focus on films grown on $Al_2O_3$ and use a film of 49 nm thick as a representative throughout this paper. Its resistance showed a superconducting transition around 13 K (Fig. 1a) and showed a down turn at around 70 K (Fig. 1a inset). The latter one is also present in the non-superconducting FeTe bulk crystals and is associated with the antiferromagnetic transition (see supplementary materials).

In order to investigate the superconducting regime below 10 K where the resistance of the film was too small to be reliably measured using charge transport (Fig. 1a), we employed the mutual inductance technique, which is sensitive to the superconductivity even in monolayer films[12,13]. The in-phase component (X) of the AC voltage on the pickup coil $V_p$ increased whereas the out-of-phase component (Y) decreased as a function of temperature below 10 K (Fig. 1b) in consistent with a diamagnetic response[14-17]. But unlike the diamagnetic response of a BCS bulk superconductor, X exhibited a broad peak and Y did not saturate down to 2 K (Fig. 1b). Such behavior has been observed in other unconventional superconducting films[14,17].

As a function of the external magnetic field applied perpendicular to the film, our sample showed strong hysteresis both from transport and from mutual inductance measurement (Fig. 2). The overall signal was symmetric about zero field and therefore we will focus our discussion on the positive field. At 2 K, resistance in the down-sweep (Fig. 2a blue) is lower than that in the up-sweep (Fig. 2a orange), leading to an enhanced critical field in the down-sweep. Similarly, Y of the mutual inductance signal was also lower in the down-sweep than in the up-sweep (Fig. 2b), leading to a peak value in the down-sweep occurring at ~ 0.2 T before the field returned to zero. The magnitude of the hysteresis as represented by the difference between up and down sweeps decreased as temperature increased (Fig. 2d). Due to the reduced X and Y signal close to Tc, no hysteresis was observable from the mutual inductance signal above 9 K (Figs. 2b-d). Nevertheless, it was still present in the resistance data up to 11 K (Fig. 2a inset).

We found that the magnetic hysteresis in the superconducting state obtained from both transport and from mutual inductance signal could be unified if we transform the mutual inductance voltage $V_p = X + iY$ into surface impedance $Z_s = R_s + i\omega L$ according to

$$V_p = i\omega I_d \int_0^\infty dx \frac{M(x)}{1 - (\frac{2xi}{\mu_0 h\omega})Z_s}$$

where $R_s$ is the surface resistance, $L$ is the inductance with $\frac{1}{L} = (\frac{2e^2}{m})n_s$ proportional to superfluid density $n_s$, $I_d$ and $\omega$ is the amplitude and frequency of the drive current, respectively, and $M(x)$ is determined by the specific geometry of the mutual inductance coils[16,17]. We used frequencies around 10 KHz for both transport and mutual inductance measurements to keep them in a similar quasi-DC regime. As we can see from Figure 3a, $R_s(T)$ can be connected with $R(T)$ from transport up to a scaling constant. They overlap in the temperature window of 10 K ~ 11 K, below which the resistance in the superconducting state is too small to be measured by charge transport and above which the surface impedance is dominated by normal inductance. By combining the two, we covered 8 decades of variation in resistance from the first sign of superconductivity at 13 K. From the imaginary part of $Z_s$, we obtained $L^{-1}$ which showed that the superfluid density almost kept a linear temperature dependence deep into the superconducting state and was unsaturated at 2 K (Fig. 3b). Such linear dependence of $n_s(T)$ has been observed in cuprate superconductors[14,18].

Applying the same method on the mutual inductance data under sweeping external magnetic field $H$ at various fixed temperatures (Fig. 2b, c and d), we were able to connect $R_s(H)$ from mutual inductance with $R(H)$ from transport (Fig. 3c and d) using the same scaling constant for connecting the temperature data (Fig. 3a). The combined $R_S(H)$ hysteresis loop (Fig. 3c) clearly illustrated two abnormal features: 1) The surface resistance is lower in the down sweep, and 2) the minimum of the surface resistance occurred at a field $H_m$ before the field crosses zero. These features are quite the opposite of what is expected from the hysteresis loop of a homogeneous superconductor with vortex pinning. Because $R_s$ is caused by motion of the free vortices, such features in hysteresis suggested the existence of different levels of pinning strength which may be a result of granularity[19].

According to the two-level critical state model, the intergranular regions have a higher penetration field $H_g$ and much stronger pinning than the grain boundaries. When the field is swept upward, flux penetrates grain-boundaries first forming Josephson vortices which causes large increase in $R_s$ due to a lower viscosity. In the down sweep, intergranular fluxons disappear first, leaving the more strongly pinned intragrain fluxons that contribute less $R_s$ per fluxon than their intergranular counterparts.

The validity of the two-level critical state model can be further seen from the variation of hysteresis with the variation of temperature and maximum applied field $H_{max}$. As we increased temperature, both the magnitude of hysteresis in $R_s$ and $H_m$ reduced and both disappeared at 12 K (Fig. 3d). $H_m$ followed a linear temperature dependence (Fig. 4a) which suggests that the penetration field of the grains and the critical current density decreases linearly with increasing temperature. At the lowest temperature, $H_m$ followed a linear dependence of $H_{max}$ when it was small and tended to a constant when $H_{max}$ was large (Fig. 4b). The turning point of these two regions is nothing but the grain's penetration field $H_g$, which can be determined by fitting $H_m$ in the low field and finding its interception point with the constant level at high field[19]. All these behaviors were in good agreement with the two-level critical state model[19].

We found that such two-level critical state invariably occurred in the superconducting FeTe films we studied. Films of same thickness with higher Tc tend to have stronger hysteresis at same temperature, which is not surprising given the temperature dependence of $H_m$ (Fig. 4a). What is a bit surprising is that $H_g$ decreased in thicker films that have very similar Tc with the others under comparison (Fig. 4c). In the two-level critical state model, $H_g$ is proportional to the grain size[19]. Our observation suggests that superconducting grains were much bigger in thinner FeTe films, which was in contrast to the roughness from topography(see supplementary materials). This points suggestively to the role of both oxygen and interface in inducing superconductivity in FeTe: thinner films allow oxygen to permeate uniformly towards the bottom layers, whose lattice is likely distorted from the FeTe bulk, and this combination facilitates larger superconducting grains to form.

It is well known that granularity strongly affects the electromagnetic properties of superconductors. Even in high quality cuprate high Tc superconductors, granularity plays a crucial part in determining the critical density [20,21], magnetization[19], superfluid density[18] and superconducting gap inhomogeneity[22]. In some cases, granularity in thin films may appreciably enhance Tc[23,24]. Our finding of the two-level critical state in the superconducting FeTe films after oxygen annealing suggests granularity may also play an essential role in inducing superconductivity from an antiferromagnetically ordered iron-chalcogenide.

**Acknowledgement**

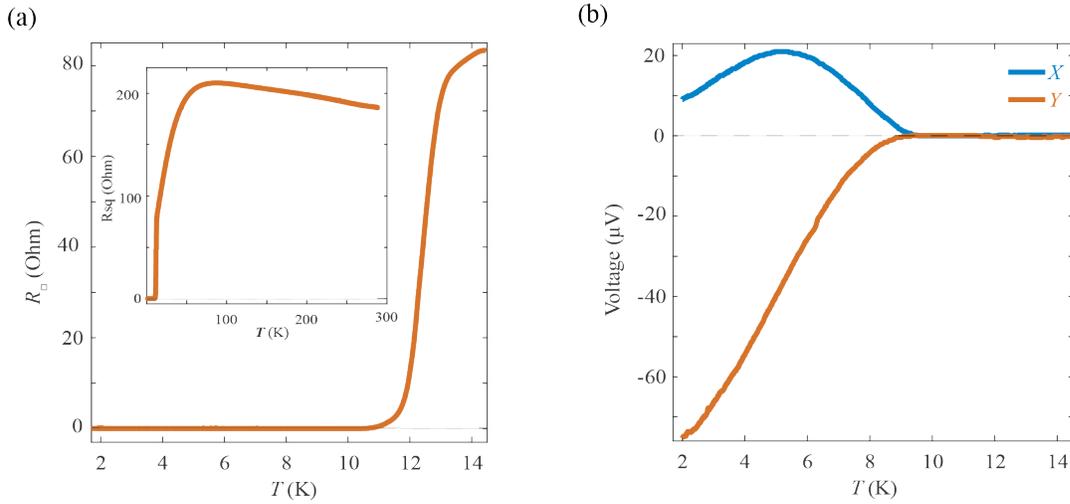

**Fig. 1 Unsaturated superconductivity in FeTe films after annealing in oxygen.** (a) The sheet resistance ($R_\square$) of a 49 nm thick FeTe sample after annealing in oxygen. Inset: resistance over a larger temperature range showing a transition around 70 K similar to the antiferromagnetic transition in the non-superconducting FeTe bulk crystals. (b) The mutual inductance signal of the same sample. Blue and orange are the in-phase (X) and out-of-phase (Y) component of the AC voltage signal on the pickup coil respectively. The data was obtained at a drive frequency of 10 KHz and drive current of 10 µA.

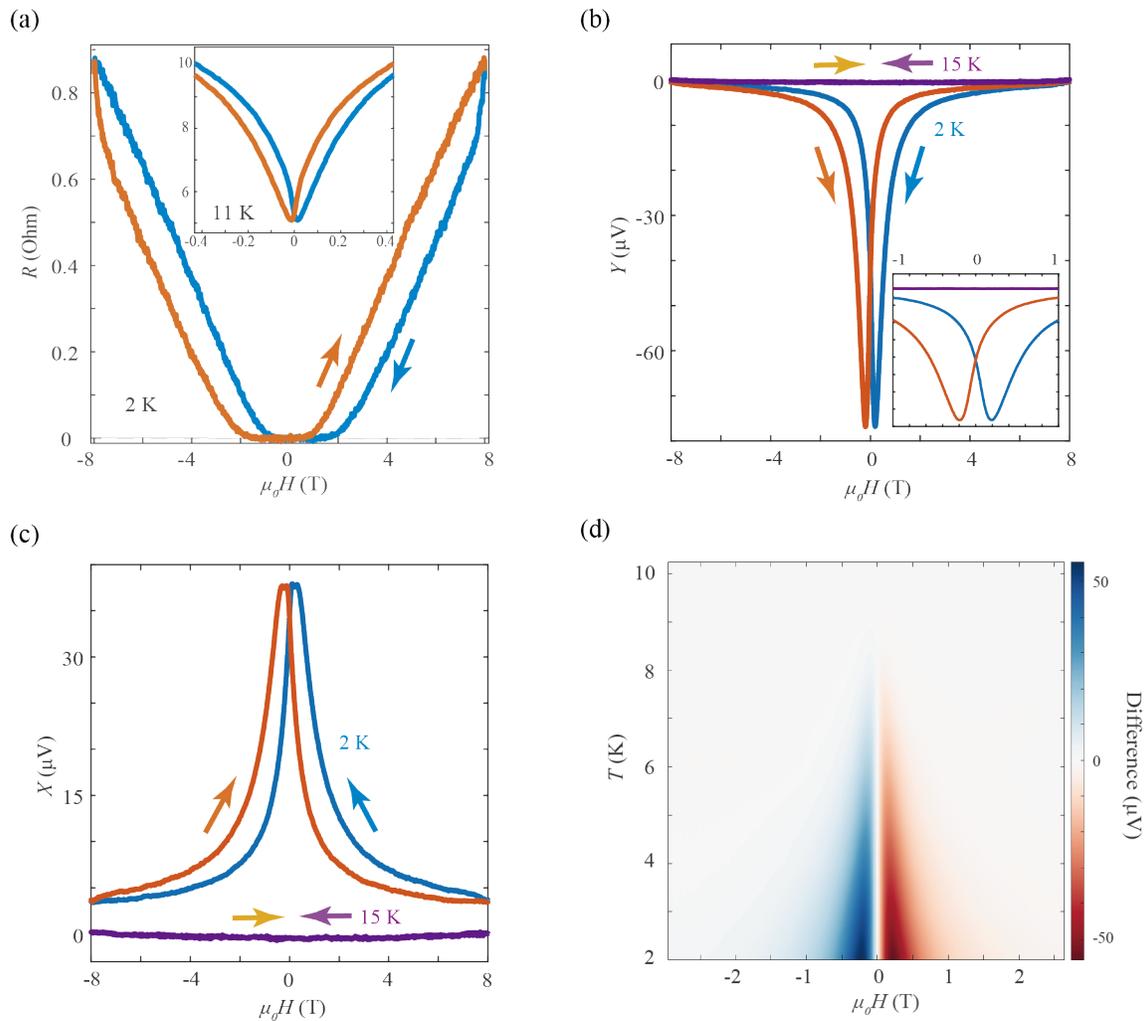

**Fig. 2 Magnetic hysteresis of the film in the superconducting state.** (a) Resistance at 2 K (inset: 11 K) showing hysteresis under up-sweep (orange) and down-sweep (blue) of the perpendicular magnetic field. The sample was cooled under nominally zero-field to 2 K. Similar hysteresis was observed from the mutual inductance measurement on the out-of-phase (b) and in-phase (c) components. The yellow and purple curves, which overlap, represent up and down sweeps, respectively, at 15 K. (d) Difference between the down and up field sweeps of out-of-phase component of the mutual inductance signal as a function of temperature.

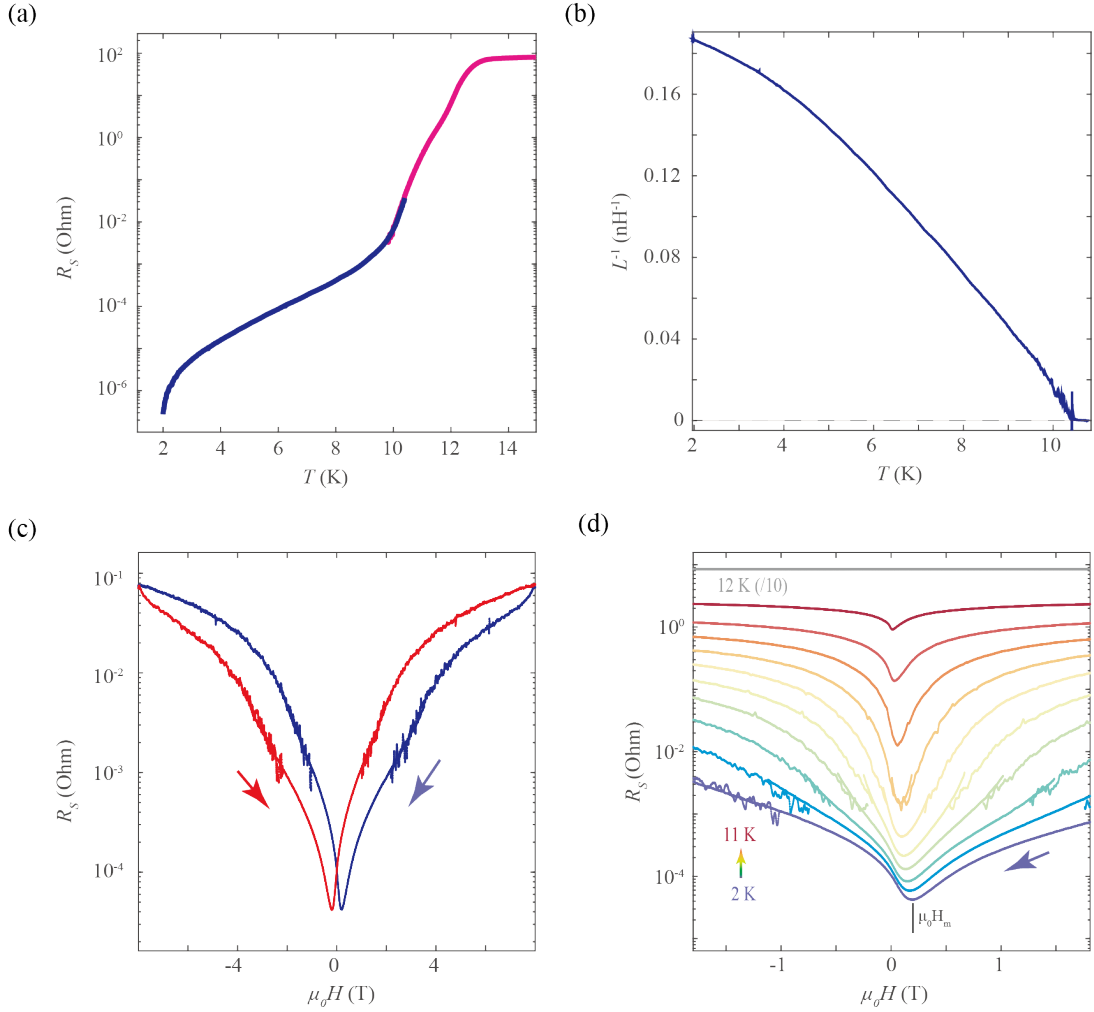

**Fig. 3 Obtaining surface impedance and the magnetic hysteresis in the surface resistance.**
(a) Surface resistance $R_S$ of the superconducting film as a function of temperature, measured by transport (magenta) and extracted from mutual inductance (purple). (b) Inverse of the surface inductance (which is proportional to the superfluid density) extracted from the mutual inductance data. See text for the details for extracting the surface impedance from mutual inductance data. (c) $R_S$ as a function of magnetic field. Again, data in the high field region was obtained from transport and that in the low field region was extracted from surface impedance. The maximum applied field ($\mu_0 H_{max}$) was 8 T. (d) Field down-sweep of $R_S$ as a function of temperature with $\mu_0 H_{max} = 8\,T$. The 12 K data (grey) was scaled down by a factor of 10. $\mu_0 H_m$ is defined as the field at which the lowest $R_S$ occurs in the down-sweep.

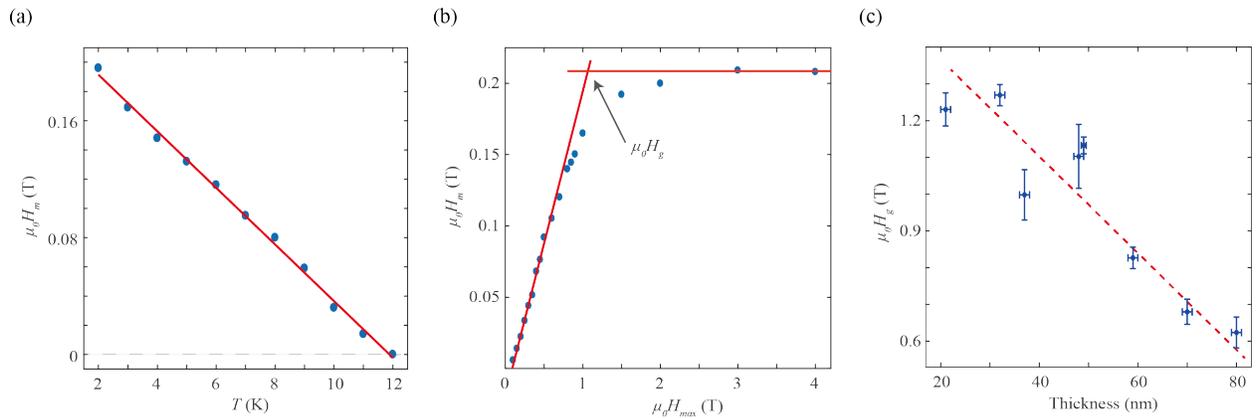

**Fig. 4 Determination of the characteristic field in the two level critical state and its film thickness dependence.** (a) The lowest field at which the lowest resistance occurs in the magnetic hysteresis $H_m$ as a function of temperature obtained from Fig. 3(d). The red line is a linear fit. (b) $H_m$ as a function of $H_{max}$ measured at 2 K. The two red lines are linear fits in the high and low field region, respectively. We can obtain the penetration field $H_g$ from the crossing point between these two lines (See text). (c) $H_g$ at 2 K as a function of film thickness. Films under comparison have similar Tc. The dashed line is a linear fit to guide the eye.